\documentclass[runningheads]{llncs}
\pdfoutput=1
\usepackage{eccv}

\usepackage{eccvabbrv}

\usepackage{graphicx}
\usepackage{booktabs}
\usepackage{wrapfig}
\usepackage{multirow}
\usepackage{footnote}
\usepackage{makecell}
\usepackage{arydshln} 
\usepackage{booktabs}
\usepackage{amssymb}
\usepackage{wrapfig} % 文字环绕宏包
\usepackage{bbding}
\usepackage{algorithm} % 算法宏包
\usepackage{algpseudocode} % 算法伪代码宏包
\usepackage[accsupp]{axessibility}  % Improves PDF readability for those with disabilities.

\usepackage{hyperref}

\usepackage{orcidlink}

\begin{document}
\pdfoutput=1
\title{Representing Topological Self-Similarity Using Fractal Feature Maps for Accurate Segmentation of Tubular Structures} 

\titlerunning{Fractal Feature Maps for Segmentation of Tubular Structures}

\author{Jiaxing Huang\inst{1,2}\orcidlink{0000-0002-0222-7925} \and
Yanfeng Zhou\inst{1,2}\orcidlink{0000-0001-5988-5331} \and
Yaoru Luo\inst{1,2}\orcidlink{0000-0002-6547-1634}\and
Guole Liu\inst{1,2}\orcidlink{0000-0002-1006-6383}\and
\\Heng Guo\inst{3,4}\orcidlink{0000-0002-4069-4743}\and
Ge Yang\inst{1,2}\orcidlink{0000-0001-6176-3130}}

\authorrunning{J.~Huang et al.}

\institute{Institute of Automation, Chinese Academy of Sciences, Beijing 100190, China \and
School of Artificial Intelligence, University of Chinese Academy of Sciences\\
\and
DAMO Academy, Alibaba Group\\
\and
Hupan Lab, Hangzhou 310023, China\\
\email{\{huangjiaxing2021, ge.yang\}@ia.ac.cn}}

\maketitle

\begin{abstract}
  Accurate segmentation of long and thin tubular structures is required in a wide variety of areas such as biology, medicine, and remote sensing. The complex topology and geometry of such structures often pose significant technical challenges. A fundamental property of such structures is their topological self-similarity, which can be quantified by fractal features such as fractal dimension (FD). In this study, we incorporate fractal features into a deep learning model by extending FD to the pixel-level using a sliding window technique. The resulting fractal feature maps (FFMs) are then incorporated as additional input to the model and additional weight in the loss function to enhance segmentation performance by utilizing the topological self-similarity. Moreover, we extend the U-Net architecture by incorporating an edge decoder and a skeleton decoder to improve boundary accuracy and skeletal continuity of segmentation, respectively. Extensive experiments on five tubular structure datasets validate the effectiveness and robustness of our approach. Furthermore, the integration of FFMs with other popular segmentation models such as HR-Net also yields performance enhancement, suggesting FFM can be incorporated as a plug-in module with different model architectures. Code and data are openly accessible at \url{https://github.com/cbmi-group/FFM-Multi-Decoder-Network}.
\end{abstract}

\footnote{This paper has been accepted by the European Conference on Computer Vision 2024.}
\section{Introduction}
\label{sec:intro}

\begin{figure}[tb]
  \centering
  \includegraphics[width=0.7\linewidth]{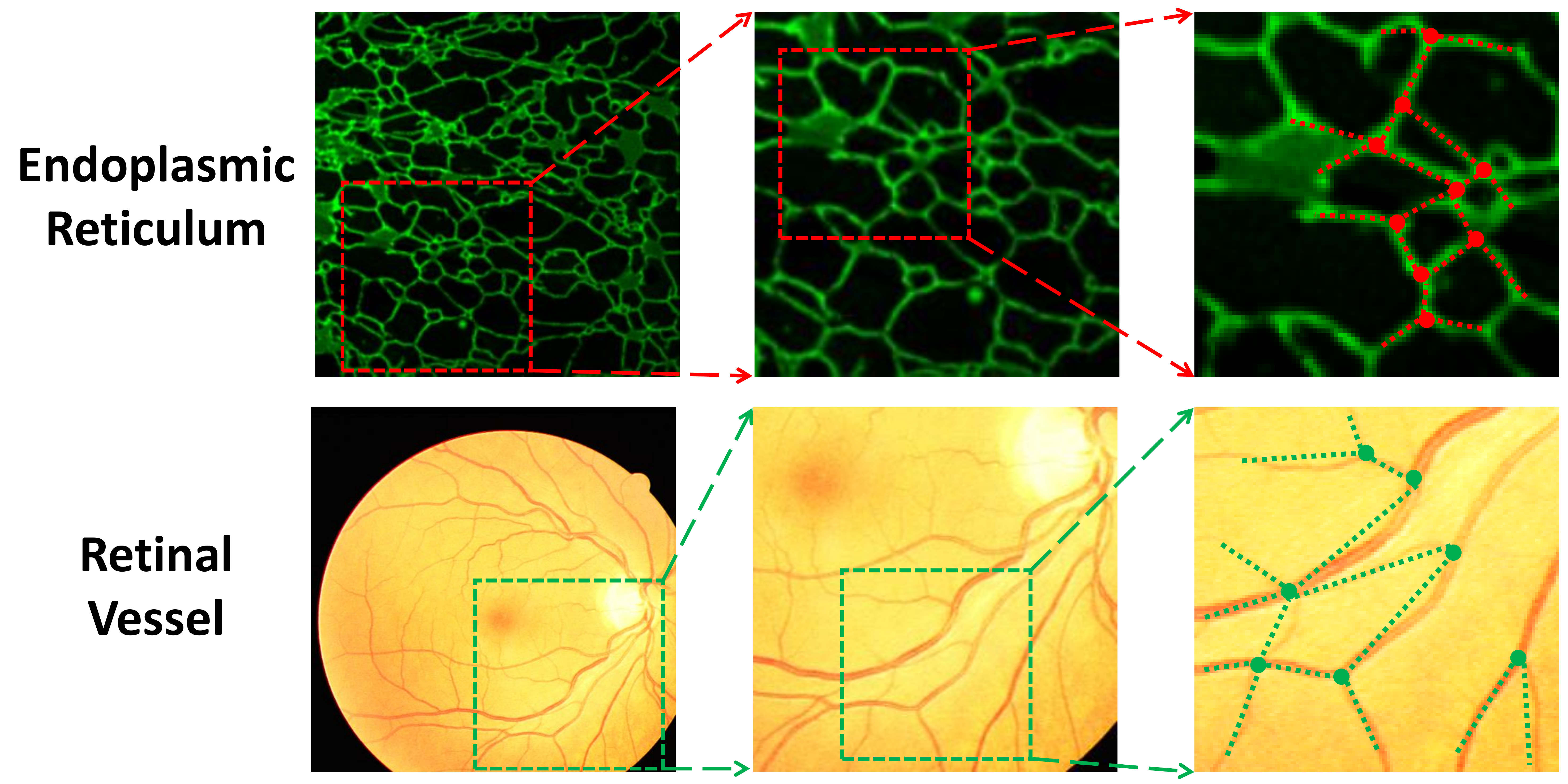}
  \caption{Topological self-similarity in tubular structures. If we consider ``one junction with multiple edges'' as a basic component, complex tubular structures have similar components at different scales. The images on the right are magnified views of the rectangular regions on the left.}
  \label{fig:1}
  \vspace{-0.4cm}
\end{figure}

Accurate segmentation of tubular structures is of significant importance across a wide variety of areas. In the area of biological research, for example, the accurate segmentation of tubular structures such as the endoplasmic reticulum (ER, \cref{fig:1}) is critical to the study of related human disease mechanisms \cite{lee2020er,hotamisligil2010endoplasmic}. In the area of clinical research, the accurate segmentation of blood vessels (\cref{fig:1}) is essential to the early diagnosis of diseases such as retinopathy and stroke \cite{mou2019cs,zhao2017automatic}. Similarly, in the area of remote sensing, the accurate extraction of roads from aerial imagery is essential for navigation and route planning \cite{laptev2000automatic}. However, accurate segmentation of tubular structures from images remains challenging due to factors such as their complex morphology and geometry, low image signal-to-noise ratio, and poor image contrast.

A wide variety of techniques have been developed for the segmentation of tubular structures. Classical methods rely on manually crafted features such as intensity, texture, and shape. For example, previous studies \cite{antiga2003computational,nain2004vessel,yim2001vessel,challoob2023distinctive} have utilized deformable shape models to fit tubular structures, leveraging their geometric properties. However, these techniques often cannot handle challenges posed by factors such as poor contrast, high noise and complex background. Deep learning methods \cite{long2015fully,ronneberger2015u,he2016deep,Zhou_2023_ICCV,kreitner2024synthetic} have revolutionized image segmentation and achieved substantial improvements in segmentation performance. Recent studies on using deep learning models for segmentation of tubular structures have focused on optimization of loss functions \cite{wang2019tubular,wang2020deep,araujo2021topological,shit2021cldice,la2023tubular} and refinement of model structures \cite{ma2020rose,hu2021topology,dong2022deu,qi2023dynamic,gupta2024topology}. However, these studies primarily aim to achieve high segmentation performance utilizing a limited input of images without providing additional information to their segmentation models. Tubular structures exhibit distinct topology and geometry that are vital for segmentation. In this study, we explore using characteristics of their structure to assist deep learning models in segmentation.

One notable characteristic of tubular structures is their topological self-similarity, namely large and complex tubular structures exhibit similar topological patterns at different scales. For example, \cref{fig:1} shows that if   ``one junction with multiple edges'' is considered a primary structural component, tubular structure entities such as the ER network and the retinal blood vessel network exhibit similar topology at both global and local scales. To quantify the topological self-similarity of tubular structures, we utilize the fractal theory, which characterizes self-similarity of intricate structures at different scales \cite{lopes2009fractal,zhuang2019application}. Central to this theory is the fractal dimension (FD), a key parameter used previously to describe the textural attributes of images. 

In addition to topology and geometry, edges and skeletons are important characteristics in defining tubular structures. In the segmentation of these structures, boundary accuracy and skeletal continuity are crucial for downstream tasks. For example, a commonly encountered problem in the segmentation of interconnected tubular structures (\cref{fig:1}) is the breakages in segmentation results due to factors such as low-contrast or blurring. 

To enhance the segmentation quality of tubular structures, we exploit the topological self-similarity as well as edge and skeletal characteristics of tubular structures. The main research contributions of this study are as follows:

1) We have developed a strategy to incorporate fractal features into deep learning networks. Specifically, we extend the fractal dimension from the image-level to the pixel-level and generate the fractal feature map (FFM), which characterizes topological self-similarity and textural complexity of each region within an image or its associated label. The FFM computed from the image is denoted as $FFM_{image}$, while the FFM derived from the label is indicated as $FFM_{label}$. Utilization of $FFM_{image}$ as an additional model input and $FFM_{label}$ as an additional loss function weight substantially enhances segmentation performance. 

2) We develop the multi-decoder network (MD-Net) by extending the U-Net \cite{ronneberger2015u} architecture with an edge decoder and a skeleton decoder. These decoders enable the model to simultaneously predict the boundaries and skeletons of image objects, in addition to the primary segmentation masks. By incorporating related constraints within the loss function, our model focuses not only on achieving accurate target segmentation but also allocates increased attention to boundary delineation and skeleton preservation, thereby enhancing the overall prediction quality.

3) We have demonstrated the versatility and robustness of FFMs. Incorporation of $FFM_{image}$ into the vanilla U-Net \cite{ronneberger2015u} and HR-Net \cite{wang2020HRNet} enhances segmentation performance, indicating that FFM can be used as a plug-in for different models.

\section{Related Work}

\subsection{Tubular Structure Segmentation}
\textbf{Classical methods} have been proposed to improve the performance of segmenting tubular structures by taking into account their geometric characteristics. Firstly, various methods utilize active contours and compute geodesics or minimal distance curves to approximate the contours of tubular structures, thereby effectively delineating their boundaries \cite{alvarez2017tracking,caselles1997geodesic}. Secondly, tree structure-based methods utilize intrinsic shape priors to assist segmentation of different tubular structures. These methods employ a bottom-up method to identify tubular objects and a top-down grouping strategy to recognize tree structures, generating corresponding shape priors \cite{bauer2010segmentation}. Lastly, various centerline-based methods such as the one developed in \cite{sironi2014multiscale} use multiscale detection strategies to accurately identify centerlines so that distance transform can be used to provide valuable information for segmentation of tubular structures. 

\textbf{Deep learning methods} have also been proposed to integrate topological and geometrical prior knowledge of tubular structures to enhance the performance of their segmentation. The integration is primarily achieved in three ways.

1) Convolutional kernel design. The popular deformable convolution \cite{dai2017deformable} and dilated convolution \cite{yu2017dilated} aim to overcome the limitations of geometric transformations in CNNs and have demonstrated exceptional performance in complex segmentation tasks. Additionally, DSC-Net utilizes dynamic snake convolution to accurately capture the distinct features of tubular structures \cite{qi2023dynamic}. By adaptively focusing on slender and winding local structures, DSC-Net achieves improved performance in capturing the intricacy of tubular structures.

2) Model architecture design. Various architecture designs have been proposed to learn the topological and geometrical features of tubular structures. In \cite{li2023global}, a global transformer and dual local attention network are employed to simultaneously capture global and local features to effectively learn the complex geometric properties of tubular structures. Dong et al. \cite{dong2022deu} propose an enhanced Deformable U-Net that exploits flexible deformable convolutional layers to better generate clear boundaries for 3D cardiac cine MRI. High-frequency components that have strong capabilities to perceive thin structures are fused in \cite{he2021thin} to enhance the performance of segmenting thin structures.

3) Loss function design. Various loss functions have been explored for the segmentation of tubular structures. In \cite{shit2021cldice}, a similarity measure centerlineDice (clDice) based on the intersection of segmentation masks and their respective skeletons is introduced. A loss function based on clDice is proposed to enable the networks to generate segmentations with more accurate connectivity information and topology preservation. Araujo et al. \cite{araujo2021topological} design a loss function based on the morphological closing operator that allows models to produce more topologically coherent masks and consistent vascular trees. Wong et al. \cite{wong2021persistent} apply a novel Persistence Diagram Loss that quantifies topological correctness of segmentation over fine-grained structures.

In this study, instead of relying on intricate designs of model architectures or loss functions, we incorporate FFMs as a model input and a loss function weight to enhance the perception of topological self-similarity and textures of images. 

\subsection{Fractal Theory and Applications}
Despite their topological and geometrical complexity, tubular structures often exhibit topological similarities across different spatial scales. This observation suggests that their complex spatial patterns can be effectively described using simple texture features. Fractal geometry provides a means to describe the irregular or fragmented shapes of natural features and other intricate objects that traditional Euclidean geometry struggles to analyze \cite{lee2010robust}. Specifically, fractal features offer the capacity to describe and characterize the topological and geometrical complexity as well as textural composition of tubular image objects. Thus, fractal geometry has been applied to image classification and segmentation tasks.

For classification, Roberto et al. \cite{roberto2021fractal} and Lin et al. \cite{lin2013automatic} utilize the fractional Brownian motion (FBM) model \cite{kaplan1994extending} to extract fractal features of images. These features are then fed into a support vector machine or a convolutional neural network (CNN) classifier to differentiate between different objects. For segmentation, the FBM model is utilized in \cite{lin2015alveolar,zhuang2019application} to extract fractal features from images. Such features are combined with classical methods such as thresholding and region growth techniques for image object segmentation. 

Although existing methods have demonstrated good performance by leveraging fractal features, there is still a gap in exploring the integration of fractal features with deep learning for segmentation tasks. The inherent self-similarity observed in tubular structures aligns well with the fractal theory. In this study, we address this gap by incorporating FFMs into the segmentation model, aiming to provide a new and reliable source of information to enhance segmentation performance.

\section{Method}
\subsection{Fractal Feature Map}
In fractal geometry, the fractal dimension (FD) provides a quantitative measure of an image's degree of self-similarity and roughness. FD can be estimated via the property of self-similarity \cite{pentland1984fractal}. Given an image $ A $, it is self-similar if $ A $ comprises $ N_r $ distinct copies of itself scaled down by a factor of $ r $. Consequently, for an image, the FD is defined as: 
\begin{equation}
     FD = \lim_{r \to 0} \frac{\log_ {} {N_r}}{\log_ {} {(1/r)}} \tag{1}
\end{equation}

Although the definition of FD based on self-similarity is simple and concise, its direct estimation becomes impractical when dealing with irregular images. To overcome this challenge and estimate the FDs of images, the box-counting method \cite{li2009improved,konatar2020box} is employed. 

\textbf{Box-counting Method:} Consider a grayscale image $ I $ with dimensions $ M \times M $, where $ L $ denotes the maximum gray level (typically $L=255$). We can model the image $ I $ as a three-dimensional space with $ (x,y) $ indicating the two-dimensional position and a third coordinate $ (z) $ denoting the gray value. Then the 3-D space is subdivided into smaller cubic regions, or ``boxes'', each with dimensions $k \times k \times h$. Here, $k$ is a given scale used be a multiple of the sidelength of a pixel in $ (x,y) $ and $h$ can be a multiple of the gray level unit in z-direction. Given $ L $ and $ k $, the value of $h$ calculated using the following formula:

\begin{equation}
    h = (L-1) \times k / M \tag{2}
\end{equation}

Given a $ k \times k $ grid located at position $(i,j)$, suppose that the minimum gray value is contained within the $m_{th}$ box ($m=\lceil minG_{i,j}/h \rceil$), and the maximum gray value within the $l_{th}$ box ($l=\lceil maxG_{i,j}/h\rceil$). The minimum number of boxes required to encompass all gray levels within the grid at $(i,j)$ is computed as:

\begin{equation}
     n_r(i,j) = l - m + 1  \tag{3}
\end{equation}

Considering all grids, the number of boxes that can cover all the patches is expressed as:
\begin{equation}
     N_r =  \sum_{\substack{i,j}}  n_r(i,j) \tag{4}
\end{equation}
where  $ r = k/M $. We can obtain a series of $ N_r $ using differing values of $ k $. Finally, the fractal dimension can be estimated from the least-squares linear fitting of $ \log_ {} {N_r} $ versus $ \log_ {} {(1/r)} $, as illustrated in Equation (1). The flow of box-counting method is summarized in the supplementary material.

\textbf{Fractal Feature Map}: Although fractal dimension and fractal features have been previously applied to classification tasks \cite{roberto2021fractal,lin2013automatic}, the inherent differences between classification and segmentation tasks preclude the direct application of image-level fractal dimension to segmentation endeavors. We extend the calculation of FD from the image-level to the pixel-level to generate the FFM of an image. As depicted in \cref{fig2}, the process begins with the utilization of a sliding window technique. Within a $5\times 5$ window, for example, the FD of this region is computed using the box-counting method. Subsequently, the window is shifted along both the horizontal and vertical directions with a step size of 1, resulting in the calculation of the FFM for the entire image. The algorithm for generating the FFM is summarized in \cref{alg:algorithm2}. 
\begin{wrapfigure}{R}{0.53\textwidth} % 
\vspace{-1.6cm}
\begin{minipage}{0.53\textwidth}
\begin{algorithm}[H]
\footnotesize
\caption{Generation of FFM.}
	\label{alg:algorithm2}
	\begin{algorithmic}
	\State \textbf{Input:} Image $I$ with size $M \times N$, window size $w\times w$.
	\State \textbf{Output:} Fractal feature map (FFM) of $I$.
	\State Assign $FFM \leftarrow \text{Float array of size } (M, N)$
	\State Assign padding size $p \leftarrow \lfloor w/2 \rfloor$
	\State $Pad_I \leftarrow \text{Linear Padding}(I, p)$
	\State\textbf{For} {$i \leftarrow 0$ \textbf{to} $M-1$ \textbf{do}}
	  \State  \qquad \textbf{For} {$j \leftarrow 0$ \textbf{to} $N-1$ \textbf{do}}
	        \State \qquad \qquad Selected region $R_{i,j}$
                \State \qquad \qquad$R_{i,j} \leftarrow Pad_I[i:i+w, j:j+w]$
	        \State \qquad \qquad $FD \leftarrow \text{Box-counting}(R_{i,j})$
	        \State \qquad \qquad $FFM[i, j] \leftarrow FD$

\State $FFM \leftarrow \text{Normalization}(FFM)$
	\end{algorithmic}
\end{algorithm}
\end{minipage}
\vspace{-1.5cm}
\end{wrapfigure}

The FFMs of the images (denoted as $FFM_{image}$) are incorporated into the segmentation model as additional input channels (Image, $FFM_{image}$), enhancing the focus on fractal structure and self-similarity after the normalization process. To mitigate the impact of image noise on the calculation of FD, we adopt the box-counting method proposed in \cite{lee2010robust} as they substitute the gray value with the standard deviation to make the method more robust.

\begin{figure}[!t]
  \centering
  \includegraphics[width=0.7\linewidth]{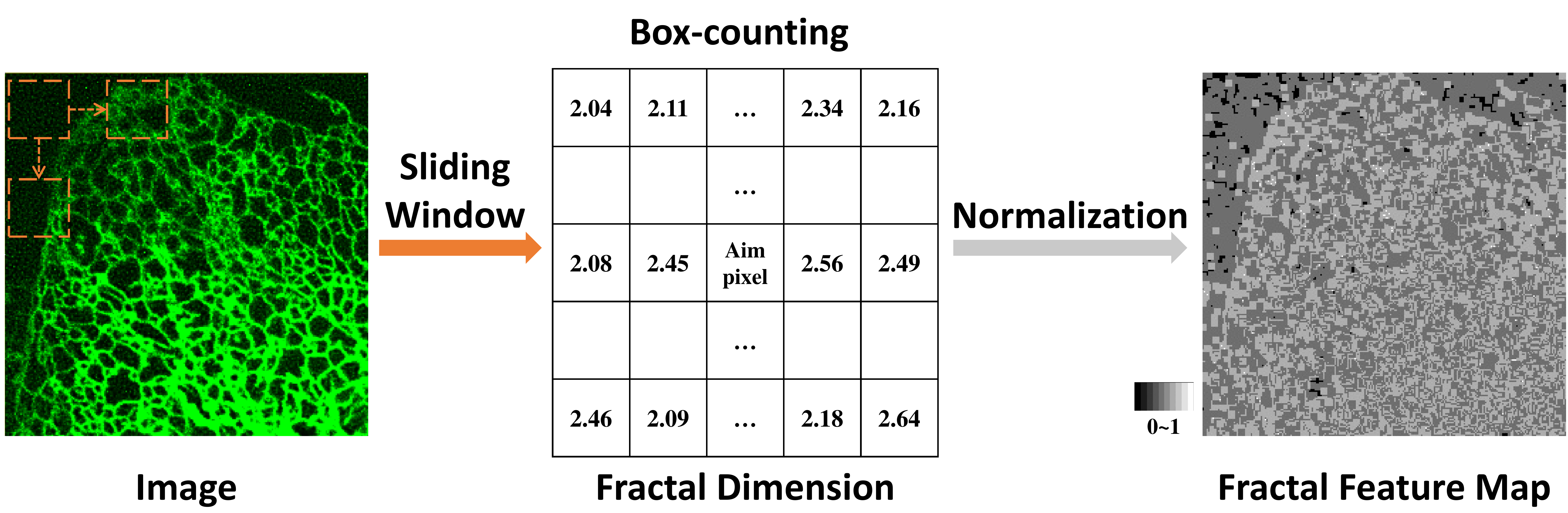}
  \caption{Workflow of computing FFM of an image.}
  \label{fig2}
  \vspace{-0.7cm}
\end{figure}

\subsection{Multi-Decoder Network}
In segmenting tubular structures, accurate boundary detection and preservation of global topological connectivity are crucial requirements. To meet these requirements, we propose a new model that we refer to as Multi-Decoder Network (MD-Net).

\begin{figure}[!b]
  \centering
\includegraphics[width=0.99\linewidth]{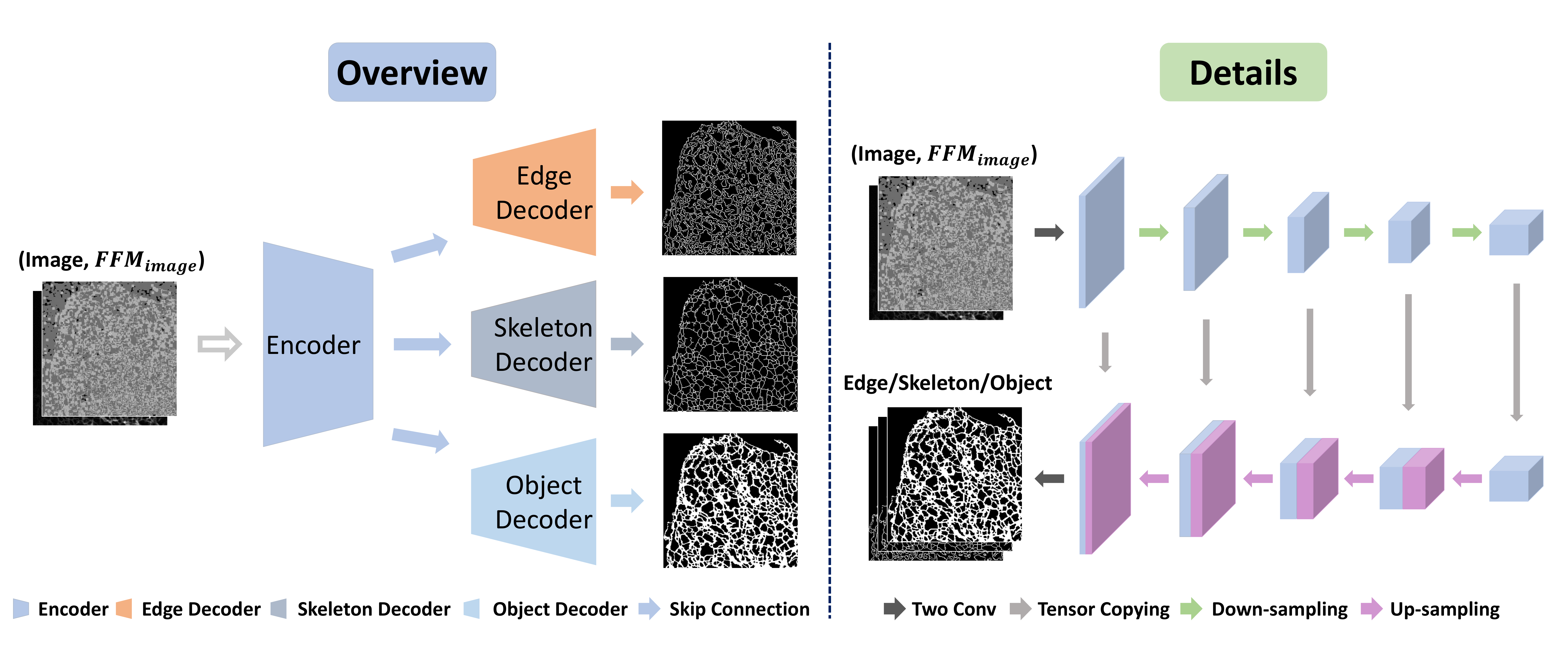}
  \caption{Overview and details of our proposed model MD-Net. The left part illustrates the overall structure of MD-Net. (Image, $FFM_{image}$) is processed as input by the Encoder, which extracts features of varying sizes from the input. These features are transmitted simultaneously to three Decoders through Skip Connection. The right part is a detailed description of the Encoder and Decoders. Each Decoder performs upsampling and concatenation operations similar to U-Net to obtain the predictions, comprising the edge, skeleton, and object.}
  \label{fig3}
    \vspace{-0.5cm}
\end{figure}

In addition to predicting the segmentation mask, we introduce an edge decoder and a skeleton decoder to generate the edge and skeleton of the tubular structure, respectively, as depicted in \cref{fig3}. When (Image, $FFM_{image}$) are fed into MD-Net, the Encoder employs convolution layers to extract a series of low-level to high-level image features. These features are simultaneously transmitted to the three Decoders by skip connections for prediction. Specifically, in the Encoder, each convolution layer involves the repeated application of two 3x3 convolutions, followed by a rectified linear unit (ReLU), and a 2x2 max pooling operation with a stride of 2 for down-sampling. As for the Decoders in MD-Net, each convolution layer includes an up-sampling of the feature map, followed by a 2x2 convolution that reduces the number of feature channels by half. Subsequently, the feature is concatenated with the corresponding feature copied from the Encoder. This concatenated feature is then processed using two 3x3 convolutions, each followed by a ReLU activation function.

During training, the ground truths for edges and skeletons are derived from the annotated masks by employing the findContours function from the OpenCV library and the skeletonization algorithms in the scikit-image library. We have visualized the boundaries and skeletons in the supplementary material to corroborate the accuracy of the extraction techniques. During the inference stage, the final structural segmentation is the output of the object decoder. Edge and skeleton predictions are only used in model training.

\textbf{Loss Function: }Given an image $I$ with $N$ pixels, the segmentation ground truth is denoted as $y$, where foreground and background pixels are labeled as 1 and 0, respectively. The prediction of the model is represented as $ \hat{y} \  \epsilon \ [0,1] $, indicating the probabilities of individual pixels being classified as foreground. To compute the loss, we utilize the differentiable Soft IOU Loss \cite{huang2019batching}, denoted as $\mathcal{L}_{object}$ in Equation (5), where $\eta$ is a smoothing coefficient.
\begin{equation}
     \mathcal{L}_{object} = \mathcal{L}_{iou} = 1- \frac{\sum_{i=1}^{N} {y_i\hat{y_i}}}{\sum_{i=1}^{N} {y_i+\hat{y_i} - y_i\hat{y_i} +\eta}}  \tag{5}
\end{equation}

The BCE loss function is applied separately to the edge and skeleton segmentation tasks, denoted as $\mathcal{L}_{edge}, \mathcal{L}_{skeleton} $. 
\begin{equation}
     \mathcal{L}_{edge}, \mathcal{L}_{skeleton}= \mathcal{L}_{bce} =
     -\frac{1}{N} \sum_{i=1}^{N} {y_i \cdot \log {(\hat{y_i})} + (1-y_i)\cdot \log {(1-\hat{y_i})}}
     \nonumber \tag{6}
\end{equation}
For MD-Net, which has three segmentation tasks, the loss function is composed as in Equation (8). The weights $\alpha,\beta$  and $\gamma$ are typically set to 1.0, 0.5 and 0.5, respectively. 
\begin{equation}
     \mathcal{L}_{global} = \alpha \mathcal{L}_{object}+  \beta \mathcal{L}_{edge} + \gamma \mathcal{L}_{skeleton} \tag{7}
\end{equation}

\subsection{Fractal Feature Constrained Loss}
The value of FD serves as an indicator of the texture complexity present in a given image region. Hence, it is advantageous to incorporate pixel-level FFM as weights of the loss function. This allows for a higher weight to be assigned to regions that exhibit greater complexity, as accurately segmenting such regions poses a greater challenge. Consequently, we calculate the FFM for each label associated with the image, denoted as $FFM_{label}$, and employ $FFM_{label}$ as pixel-level weights within the loss function to guide the convergence of the model. Considering the specificity of the edge decoder and skeleton decoder, we have thus restricted the application of $FFM_{label}$ exclusively to the object decoder. The Constrained Loss of MD-Net is as follows:
\begin{equation}
     \mathcal{L}_{constrained} = \alpha \mathcal{L}_{object}\cdot FFM_{label}+  \beta \mathcal{L}_{edge} + \gamma \mathcal{L}_{skeleton} \tag{8}
\end{equation}

\section{Experiments}

\subsection{Datasets}
We evaluate the FFM and MD-Net using five publicly available tubular datasets including the endoplasmic reticulum (ER) and mitochondrial (MITO) \cite{luo2020ermito, guo2021segmentation} organelle network datasets, the Retinal OCT-Angiography vessel Segmentation (ROSE) and Structured Analysis of the Retina (STARE) \cite{ma2020rose, hoover2003locating} retinal vessel datasets, and the Massachusetts Roads (ROAD) \cite{mnih2013machine} remote sensing dataset. By selecting these datasets, our aim is to comprehensively evaluate the performance of our methods across different types of structures and domains, providing a robust assessment of their capabilities. To further examine the influence of FFM, we included the non-tubular dataset NUCLEUS \cite{caicedo2019nucleus}. 

All images in ER, MITO, STARE, and NUCLEUS are randomly sampled patches with size of $256\times256$ for training and sliding window with size of $256\times256$ for validation and testing. A crop size of $256\times256$ is used randomly during training and regularly during testing for the ROAD dataset as in \cite{qi2023dynamic}. Horizontal and vertical flipping as well as $90^{\circ}$/$180^{\circ}$/$270^{\circ}$ rotation are used for data augmentation. In addition, data augmentation for ROSE is conducted by rotation of an angle from $-10^{\circ}$ to $10^{\circ}$ as described in \cite{ma2020rose}. Detailed partitions of the six datasets and experimental setup are shown in the supplementary material. 
\subsection{Performance Comparison}
We select several segmentation networks for comparison, including the vanilla U-Net \cite{ronneberger2015u}, U-Net++ \cite{zhou2018unet++}, nnU-Net \cite{isensee2021nnu}, and HR-Net \cite{wang2020HRNet}. Additionally, we compare our methods with SOTA models for diverse datasets. For the task of retinal vessel segmentation, we choose three-stage \cite{yan2018three}, OCTA-Net \cite{ma2020rose}, GT-DLA \cite{li2023global}, and AF-Net \cite{shi2023affinity}, as these networks are tailored for this task. Furthermore, we compare the performance of MD-Net in the ROAD dataset with DCU-Net \cite{yang2022dcu}, Dconn-Net \cite{yang2023directional}, and the DSC-Net \cite{qi2023dynamic}, a tubular structure segmentation network with specially designed loss $\mathcal{L}_{tc}$. To assess the generalization of FFM, we apply $FFM_{image}$ to both U-Net and HR-Net. All models are trained on the same dataset with the same hyperparameter settings.

\subsection{Evaluation Metrics}
The models are assessed using three types of metrics: volumetric, topology, and distance.

1. Volumetric scores, including intersection-over-union (IoU), accuracy (ACC), centerlineDice (clDice) \cite{shit2021cldice}, and AUC. These metrics quantify the overlap and agreement between the prediction and ground truth.

2. Topology scores. To evaluate the topological correctness of the segmentations, we calculate the Betti Error \cite{hu2019topology} $\beta$ for the 
 sum of Betti Numbers $\beta_0$ and $\beta_1$. In the evaluation of ROSE and STARE, $\beta$ Error represents the $\beta_1$ only. 

3. Distance score: Hausdorff Distance (HD) \cite{taha2015metrics} quantifies the accuracy of the boundary delineation and provides information about the spatial closeness of the predicted and ground truth boundaries.

Since the nucleus is oval in shape and not tubular, no Betti Error and clDice are used in segmentation performance evaluation and no skeletal decoder is used in the MD-Net. In this study, the evaluation metrics IoU, ACC, AUC, and clDice, are expressed as percentages (\%). The Hausdorff Distance (HD) is measured in pixels (px). All performance metrics are calculated for each image and averaged. 

\subsection{Configurations}
All models are implemented using PyTorch (version 1.12.0) and executed on 4 NVIDIA 3090 GPU cards. During the training phase, we employed the SGD optimizer with an initial learning rate of 0.05. To dynamically adjust the learning rate as the training progressed, we incorporated warm-up and exponential decay techniques. For all datasets, a fixed batch size of 32 is employed, ensuring consistency in the training process. Additionally, to prevent overfitting, a regularization weight of 0.0005 is applied. The entire training process spanned 50 epochs, providing sufficient iterations for model convergence and learning.
\begin{table}[!t]
\caption{Bold numbers indicate the best performance and italicized numbers represent the next best performance. The result marked with a star (*) represents the model input replaced by (Image) with(Image, $FFM_{image}$). $\mathcal{L}_{tc}$ in (b) is Wasserstein-distance-based TCLoss proposed in \cite{qi2023dynamic}. $\mathcal{L}_{constrained}$ represents the integration of  $FFM_{label}$ as a loss function weight, which leads to further improvement.}
    \label{tab2}
    \centering
        \begin{subtable}[t]{1\linewidth}
        \resizebox{\textwidth}{!}{
            \begin{tabular}{c|c|cccccc|cccccc}
      \hline

% \cmidrule{3-8}    
\multirow{2}[1]{*}{\textbf{Model}} & \multirow{2}[1]{*}{\textbf{Loss}} & \textbf{IoU$\uparrow$}   & \textbf{clDice$\uparrow$} & \textbf{ACC$\uparrow$}& \textbf{AUC$\uparrow$}   &\textbf{$\beta$  Error$\downarrow$ }    &\textbf{HD$\downarrow$} & \textbf{IoU$\uparrow$}   & \textbf{clDice$\uparrow$} & \textbf{ACC$\uparrow$}& \textbf{AUC$\uparrow$}   &\textbf{$\beta$  Error$\downarrow$ }    &\textbf{HD$\downarrow$} \\
\cline{3-14}          &       & \multicolumn{6}{c|}{\textbf{ER}}              & \multicolumn{6}{c}{\textbf{MITO}} \\
    \hline
    % SAM\cite{kirillov2023segment}   & $\mathcal{L}_{iou}$   & 39.66  & 58.70  & 80.19  & 69.88  & 1309  & 9.65  & 53.57  & 70.30  & 94.52  & 85.06  & 400  & 6.11  \\
    U-Net++\cite{zhou2018unet++} & $\mathcal{L}_{iou}$   & 75.02  & 94.67  & 91.04  & 97.31  & 26.02  & 7.04  & 79.70  & 97.30  & 98.19  & 99.59  & \textit{2.30} & 4.24  \\
    nnU-Net\cite{isensee2021nnu} & $\mathcal{L}_{iou}$   & 73.41  & 94.51  & 91.29  & 87.95  & 25.15  & 6.96  & 79.24  & 97.16  & 98.07  & 94.62  & 3.40  & 4.39  \\
    DSC-Net\cite{qi2023dynamic} & $\mathcal{L}_{iou}$   &  75.51 & 94.44  & 91.56 & 97.09  & 34.51  & 6.92  & 80.32  &  97.16 &  98.16 & \textit{99.65} & 2.70  & 4.36  \\
     Dconn-Net\cite{yang2023directional} & $\mathcal{L}_{iou}$   &  75.95 & 95.24  & 91.59 & 96.60  & 20.31  & 6.97  & 79.82  &  97.20 &  98.03 & 99.54 & 2.90  & 4.32  \\
      AF-Net\cite{shi2023affinity} & $\mathcal{L}_{iou}$   &  76.01 & 94.58  & 91.81 & 97.37  & 29.92  & 6.79  & 80.36  &  96.94 &  98.08 & 99.64 & 2.60  & 4.18  \\
       GT-DLA\cite{li2023global} & $\mathcal{L}_{iou}$   &  75.89 & 94.92  & 91.82 & 97.31  & 20.26  & 6.84  & 80.25  &  97.16 &  98.19 & 99.60 & 2.70  & 4.21  \\
    U-Net\cite{ronneberger2015u} & $\mathcal{L}_{iou}$   & 75.44  & 94.63  & 91.82  & 97.35  & 28.72  & 6.87  & 79.77  & 96.91  & 98.07  & 99.61  & 2.80  & 4.56  \\
    HR-Net\cite{wang2020HRNet} & $\mathcal{L}_{iou}$   & 75.83  & 95.08  & 91.71  & 97.36  & 22.57  & 6.90  & 79.63  & 97.27  & 98.14  & 99.62  & 2.90  & 4.25  \\
    \hdashline
    U-Net* & $\mathcal{L}_{iou}$   & 76.59  & 95.43  & 92.02  & 97.56  & 20.78  & 6.81  & 80.71  & 97.42  & 98.21  & 99.63  & 2.70  & 4.27  \\
    HR-Net* & $\mathcal{L}_{iou}$   & 76.43  & 95.47  & 91.95  & 97.55  & 20.52  & 6.83  & 80.62  & 97.29  & 98.17  & 99.62  & 3.30  & 4.18  \\
    MD-Net* & $\mathcal{L}_{global}$   & \textit{77.01} & \textbf{95.78} & \textit{92.06} & \textit{97.59} & \textbf{19.10} & \textit{6.77} & \textit{81.11} & \textbf{97.72} & \textit{98.25} & \textbf{99.66} & \textbf{2.20} & \textit{4.16} \\
    MD-Net* & $\mathcal{L}_{constrained}$ & \textbf{77.09} & \textit{95.74} & \textbf{92.14} & \textbf{97.65} & \textit{19.52} & \textbf{6.72} & \textbf{81.18} & \textit{97.61} & \textbf{98.26} & \textit{99.65} & 2.80  & \textbf{4.14} \\
    \hline
          &       & \multicolumn{6}{c|}{\textbf{ROSE}}            & \multicolumn{6}{c}{\textbf{STARE}} \\
    \hline
    % DU-Net\cite{jin2019dunet} & $\mathcal{L}_{iou}$   & 60.06  & 67.39  & 91.18  & 93.34  & 8.10  & 7.56  & 65.27  & 76.16  & 94.01  & 95.38  & 3.35  & 6.96  \\
    % CS-Net\cite{mou2019cs} & $\mathcal{L}_{iou}$   & 61.39  & 67.86  & 91.52  & 93.92  & 7.56  & 7.48  & 65.10  & 75.37  & 94.55  & 96.13  & 3.43  & 6.50  \\
    three-stage\cite{yan2018three} & $\mathcal{L}_{iou}$   & 62.11  & 67.07  & 91.79  & 93.41  & 9.13  & 7.28  & 66.83 & 76.71  & 94.86  & 96.43  & 3.45  & 6.57  \\
    OCTA-Net\cite{ma2020rose} & $\mathcal{L}_{iou}$   & 63.27  & 66.32  & 92.34  & 94.53 & 10.22 & 7.11 & 65.37 & 76.74  & 94.62  & 95.44  & 3.22  & 6.47  \\
    DSC-Net\cite{qi2023dynamic} & $\mathcal{L}_{iou}$   & 63.15 &  67.59 & 91.98 &  94.43 & 7.00  & 7.36  & 67.01  & 76.26  & 94.97  & 96.50 & 3.15  & 6.31  \\
    Dconn-Net\cite{yang2023directional} & $\mathcal{L}_{iou}$   &  60.37 & 64.51  & 88.57 & 93.26  & 5.73  & 8.18  & 66.14  & 76.33 & 94.78 & 96.65 & 3.77  & 6.48  \\
      AF-Net\cite{shi2023affinity} & $\mathcal{L}_{iou}$   &  62.33 & 67.05  & 91.73 & 92.55  & 8.88  & 7.32  & 67.27  &  77.17 &  94.91 & 96.47 & 3.18  & 6.49  \\
       GT-DLA\cite{li2023global} & $\mathcal{L}_{iou}$   &  63.57 & 66.31  & 92.27 & 94.18  & 9.89  & 7.16 & 67.12 & 76.32 & 94.72 &  95.79  & 3.12  & 6.63  \\
    U-Net\cite{ronneberger2015u} & $\mathcal{L}_{iou}$   & 61.52  & 67.53  & 91.33  & 94.04  & 8.22  & 7.42  & 66.15 & 76.05  & 94.68  & 96.39  & 3.22  & 6.67  \\
    HR-Net\cite{wang2020HRNet} & $\mathcal{L}_{iou}$   & 63.09  & 67.71  & 92.08  & 94.38  & 8.33  & 7.26  & 67.03 & 76.92  & 94.92  & 96.29  & 2.73  & 6.53  \\
    \hdashline
    U-Net* & $\mathcal{L}_{iou}$   & 64.07  & 67.95  & 92.25  & 94.47  & 7.88  & \textit{7.10} & 68.07  & 77.39 & 95.15  & 96.63  & 2.77  & 6.33  \\
    HR-Net* & $\mathcal{L}_{iou}$   & 64.68 & 69.42 & 92.14  & \textit{94.77} & 6.01 & 7.21  & 68.22  & 77.49 & 95.13  & 96.52  & 3.21  & \textit{6.32}  \\
    MD-Net* & $\mathcal{L}_{global}$   & \textit{65.07}  & \textbf{69.78}  & \textit{92.36} & \textbf{94.88}  & \textbf{4.22} & \textit{7.10}  & \textit{68.49} & \textit{77.79} & \textit{95.20} & \textit{96.91} & \textbf{2.57} & \textbf{6.29} \\
    MD-Net* & $\mathcal{L}_{constrained}$ & \textbf{65.19} & \textit{69.58} & \textbf{92.42} & \textbf{94.88} & \textit{4.89}  & \textbf{7.06}  & \textbf{68.73} & \textbf{78.20} & \textbf{95.26} & \textbf{96.99} & \textit{2.60} & 6.34 \\
    \hline

    \end{tabular}}
        \caption{Quantitative experimental results for the ER, MITO, ROSE, and STARE dataset.}
        \end{subtable}
        \begin{subtable}[t]{0.56\linewidth}
           \resizebox{\textwidth}{!}{
 \begin{tabular}{c|c|cccc|cc}
    \hline
    \textbf{Model} & \textbf{Loss} & \textbf{IoU$\uparrow$}   & \textbf{clDice$\uparrow$} & \textbf{ACC$\uparrow$}& \textbf{AUC$\uparrow$}   &\textbf{$\beta$  Error$\downarrow$ }    &\textbf{HD$\downarrow$} \\
    \hline
    DCU-Net\cite{yang2022dcu} & $\mathcal{L}_{ce}$    & 62.92  & 86.98  & 98.03  & 98.34  & 2.56  & 8.04  \\
    % Transunet\cite{chen2021transunet} & $\mathcal{L}_{ce}$    & 61.11  & 86.04  & 97.97  & 98.23  & 2.68  & 8.11  \\
    DSC-Net\cite{qi2023dynamic} & $\mathcal{L}_{ce}$    & 63.99  & 87.74  & 98.05  & 98.39  & 2.56  & 7.96  \\
    DSC-Net\cite{qi2023dynamic} & $\mathcal{L}_{tc}$    & 64.22  & 87.64  & 98.05  & 98.46  & 2.45  & 7.34  \\
    Dconn-Net\cite{yang2023directional}  & $\mathcal{L}_{tc}$    & 65.14  & 87.61  & 97.30  & 98.29  & 2.25  & 6.97 \\
   AF-Net\cite{shi2023affinity}  & $\mathcal{L}_{tc}$    & 65.03  & 87.27  & 97.41  & 98.38  & 2.21 & 7.29  \\
    GT-DLA\cite{li2023global} & $\mathcal{L}_{tc}$    & 64.76  & 86.80  & 97.43  & 97.94  & 2.38  & \textit{6.94}  \\
     U-Net\cite{ronneberger2015u} & $\mathcal{L}_{ce}$    & 62.47  & 86.87  & 97.97  & 98.29  & 2.61  & 8.11  \\
    HR-Net\cite{wang2020HRNet} & $\mathcal{L}_{ce}$    & 62.89  & 86.74  & 98.06  & 98.39  & 2.58  & 8.06  \\
    \hdashline
    U-Net* & $\mathcal{L}_{ce}$    & 65.74  & 87.74  & 98.38  & 98.72  & 2.48  & 6.98  \\
    HR-Net* & $\mathcal{L}_{ce}$    & 65.60  & 87.61  & 98.32  & 98.79  & 2.31  & 6.98  \\
    MD-Net* & $\mathcal{L}_{global}$    & \textit{66.07} & \textbf{88.08} & \textbf{98.43} & \textbf{98.80} & \textit{2.19} & \textbf{6.92} \\
    MD-Net* & $\mathcal{L}_{constrained}$ & \textbf{66.16} & \textit{88.07} & \textit{98.42} & \textit{98.77} & \textbf{2.06} & 6.97 \\
    \hline

    \end{tabular}%
    }
    \caption{Results for ROAD dataset.}
    \end{subtable}
    \begin{subtable}[t]{0.42\linewidth}
           \resizebox{\textwidth}{!}{
  \begin{tabular}{c|c|ccc|c}

    \hline
     \textbf{Model} & \textbf{Loss} & \textbf{IoU$\uparrow$}  & \textbf{ACC$\uparrow$}& \textbf{AUC$\uparrow$} &\textbf{HD$\downarrow$} \\
    \hline
    CS-Net\cite{mou2019cs}   & $\mathcal{L}_{iou}$   & 77.96  & 98.12  & 98.58  & 4.13  \\
    Transunet\cite{chen2021transunet} & $\mathcal{L}_{iou}$    & 78.21  & 98.34  & 98.64  & 4.32\\
    DU-Net\cite{jin2019dunet}& $\mathcal{L}_{iou}$    & 78.09  & 98.46  & 98.94  & 4.19\\
    U-Net++\cite{zhou2018unet++}  & $\mathcal{L}_{iou}$   & 78.33  & \textbf{98.59} & \textbf{99.69} & 4.09  \\
    nnU-Net\cite{isensee2021nnu} & $\mathcal{L}_{iou}$   & 78.28  & 98.57  & 94.93  & 4.09  \\
    OCTA-Net\cite{ma2020rose} & $\mathcal{L}_{iou}$   & 75.30  & 98.34  & 99.26  & 4.43  \\
    U-Net\cite{ronneberger2015u}  & $\mathcal{L}_{iou}$   & 78.02  & 98.54  & 99.65  & 4.23  \\
    HR-Net\cite{wang2020HRNet} & $\mathcal{L}_{iou}$   & 77.70  & 98.50  & 99.65  & 4.15  \\
    \hdashline
    U-Net* & $\mathcal{L}_{iou}$   & 78.46  & \textit{98.58} & \textit{99.68} & 4.15  \\
    HR-Net* & $\mathcal{L}_{iou}$   & 77.83  & 98.53  & 99.66  & 4.15  \\
    \textit{MD-Net*} & $\mathcal{L}_{global}$   & \textit{78.66} & \textbf{98.59} & \textit{99.68} & \textit{4.08} \\
    \textit{MD-Net*} & $\mathcal{L}_{constrained}$ & \textbf{78.77} & \textbf{98.59} & 99.66  & \textbf{4.06} \\
    \hline
    \end{tabular}%
    }
    \caption{Results for NUCLEUS dataset.}
    \end{subtable}
 \vspace{-0.8cm}
\end{table}
\subsection{Quantitative Evaluation}
Based on the results presented in \cref{tab2}, several conclusions can be drawn.

\textbf{Performance of MD-Net:} Our proposed model, MD-Net, demonstrates superior performance in terms of segmentation accuracy and topological continuity compared to the other models. This superiority is observed across five tubular datasets as well as one non-tubular dataset. We have also performed t-test to check whether improvements of our methods over competing methods in performance are statistically significant. Results are provided in supplementary.

In tubular datasets, MD-Net outperforms other methods in terms of segmentation results. Specifically, in the ER, MITO, ROSE, STARE, and ROAD datasets, MD-Net achieves improvements in IoU of $1.08\%$, $0.82\%$, $1.62\%$, $1.46\%$, and $1.02\%$, respectively, compared to existing SOTA methods. Furthermore, when considering topological continuity and boundary extraction, MD-Net demonstrates the best performance with the minimum $\beta$ error and Hausdorff Distance. These results highlight the ability of MD-Net, equipped with an edge decoder and a skeleton decoder, to effectively capture edge and skeleton features of thin tubular structures, resulting in more accurate and continuous topology segmentation outcomes. 

Even in the non-tubular dataset NUCLEUS, MD-Net achieves the best segmentation results with an IoU of $78.77\%$, ACC of $98.59\%$, and HD of $4.06$. This demonstrates the effectiveness of MD-Net and FFM in datasets with simpler structures.

\textbf{Generalization of FFM:} $FFM_{image}$ is incorporated into U-Net and HR-Net as an additional input channel without modifying the training parameters. The results in \cref{tab2} indicate that the inclusion of $FFM_{image}$ leads to improved segmentation performance for both U-Net and HR-Net.

For U-Net, the segmentation performance (IoU) shows improvements of $1.14\%$, $0.94\%$, $2.55\%$, $1.92\%$, $3.27\%$, and $0.44\%$ on the ER, MITO, ROSE, STARE, ROAD, and NUCLEUS datasets with the assistance of $FFM_{image}$, respectively. Similarly, for HR-Net, the IoU improves by $0.60\%$, $0.99\%$, $1.59\%$, $1.19\%$, $2.71\%$, and $0.13\%$ on the same datasets, respectively. On average, $FFM_{image}$ brings an improvement of $1.96\%$ and $1.42\%$ in IoU for U-Net and HR-Net, respectively, in the segmentation task of tubular datasets.

In conclusion, FFM demonstrates notable generalization capabilities, leading to enhanced performance across various models. Particularly in datasets with intricate structures, FFM consistently outperforms the original models in segmentation tasks. This improvement can be attributed to the inherent characteristics of FD involved in FFM.

\textbf{$FFM_{label}$ and $\mathcal{L}_{global}$ :} Fractal Dimension is a significant metric used to quantify the complexity of image textures. In the context of model training, we leverage the $FFMs_{label}$ as weights for the loss function. This method enables us to guide the model's focus towards regions with higher complexity, as indicated by a higher FD. Through extensive experimentation, we have observed that incorporating FFM-based weights into the loss function yields improvements in both the training performance and efficiency of the model, as depicted in \cref{tab2}. These findings highlight the tangible benefits of including $FFM_{label}$ as a weight parameter in optimizing the training process.

\subsection{Qualitative Evaluation}
\begin{figure}[!t]
  \centering
\includegraphics[width=0.75\linewidth]{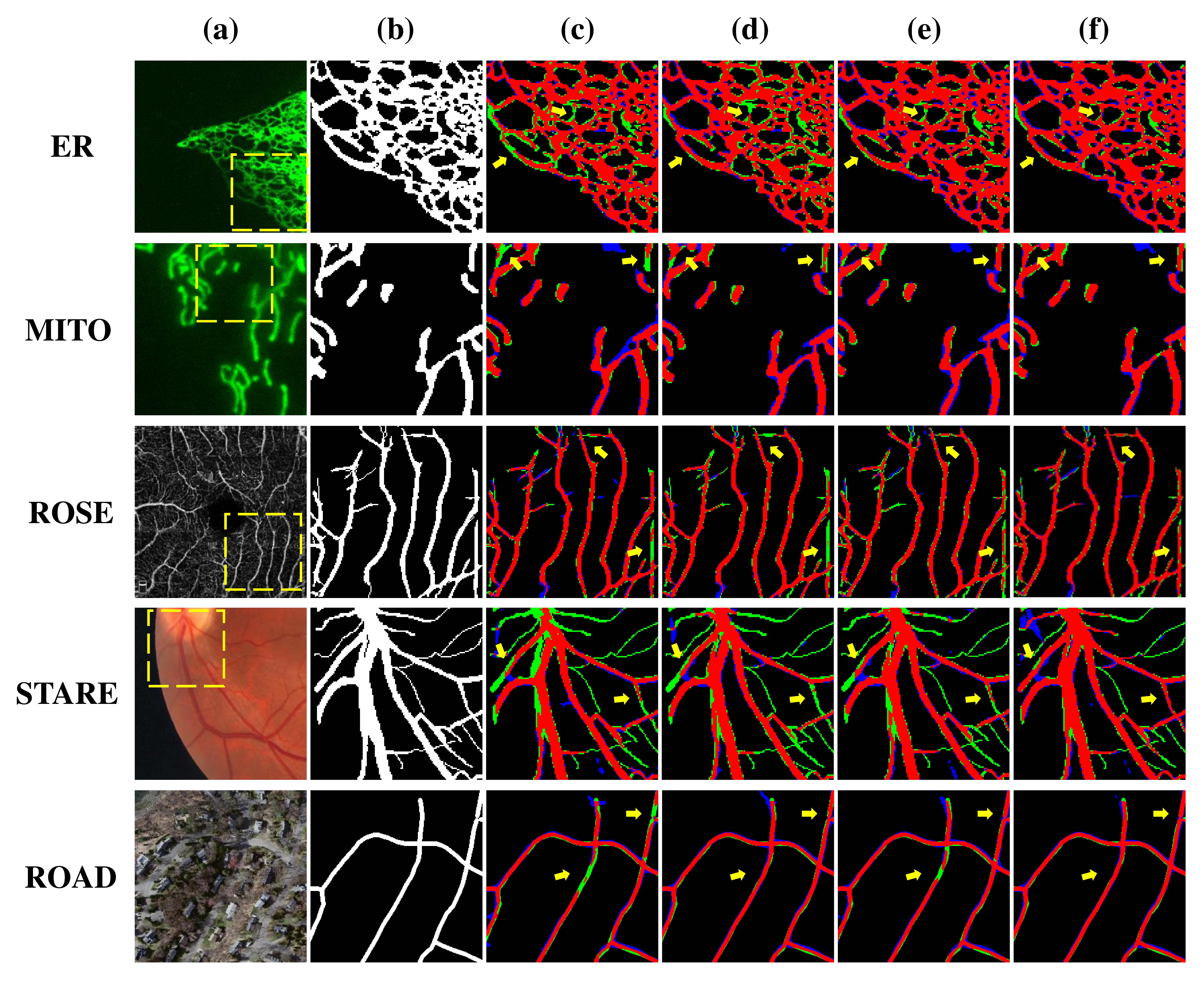}
  \caption{Comparison of segmentation results. (a) Image. (b) Label. (c) Results of U-Net. (d) Results of existing SOTA approaches. From top to bottom, it's AF-Net, AF-Net, GT-DLA, AF-Net, and Dconn-Net. (e) Results of U-Net*. (f) Results of MD-Net*. Red: true positive. Green: false negative. Blue: false positive. In the ER, MITO, ROSE, STAR, and ROAD rows, columns (b) through (f) display the magnified results of the regions demarcated by yellow squares in the corresponding images of column (a).}
  \label{fig4}
  \vspace{-0.5cm}
\end{figure}

To facilitate a more comprehensive comparison of the experimental outcomes across various models, we employed visualizations to present the segmentation results, as depicted in \cref{fig4}. The visual analysis provides a clear and intuitive representation of the model's performance. The observed results validate that the integration of FFM leads to enhanced segmentation outcomes in terms of both edge accuracy and topological continuity. More visualization results can be found in the supplementary material.

\section{Ablation Study}
\subsection{Effectiveness of Fractal Feature Maps}
To ascertain that the observed performance improvement of the model is attributed to the FFM and not the increase of input channels, we replace the $FFM_{image}$ with the image, thereby transforming the input (image, $FFM_{image}$) to (image, image). Additionally, we computed the image’s Hurst feature (HF, classical fractal analysis), mean feature (MF, directional analysis), and contrast feature (CF, directional analysis) as introduced by \cite{kaplan1999extended}. These features replaced the $FFM_{image}$ as input and were trained and evaluated on U-Net and MD-Net. Experimental results demonstrated the advantages of the $FFM_{image}$ as shown in \cref{tab3}. This finding indicates that the inclusion of the FFM plays an effective role in achieving better performance in the segmentation task.

\subsection{Robustness of Fractal Feature Maps}
The computation of the FFM is influenced by the selection of window size and step size. We used different window and step sizes to generate different $FFMs_{image}$ and tested their performance. The results in \cref{tab4} demonstrate the robustness of FFM. When used for training the U-Net and MD-Net, the $FFMs_{image}$ calculated using different window sizes consistently yield results with a variation range of $~0.3\%$ on the ER and STARE datasets. Similarly, applying different steps during the calculation of FFMs does not lead to significant changes in performance. It should be noted that the efficiency of computing can be substantially improved by increasing the step size. 

The numerical results in \cref{tab3} and \cref{tab4} are calculated by volumetric score IoU. Results with more metrics can be found in the supplementary material.

\begin{table}[b]
 \vspace{-0.5cm}
\caption{Ablation study of FFM in U-Net and MD-Net. The values within the parentheses in the (image,$FFM_{image}$) column represent the percentage improvement of IoU compared to the baseline results listed in the (image) column.}
\label{tab3}
  \centering
 \resizebox{0.76\textwidth}{!}{
    \begin{tabular}{c|c|cccccc}
    \hline
    \multirow{2}[1]{*}{\textbf{Model}} & \multirow{2}[1]{*}{\textbf{Dataset}} & \multicolumn{6}{c}{\textbf{Input}} \\
\cline{3-8}          &       & (image) & (image,image) & (image,HF) & (image,MF) & (image,CF) & (image,$FFM_{image}$) \\
    \hline
    \multirow{5}[1]{*}{U-Net} & ER    & 75.44 & 75.61 & \textit{75.89} & 75.86 & 75.10  & \textbf{76.59}(+1.14\%) \\
          & MITO  & 79.77 & 80.28 & \textit{80.51} & 80.23 & 80.38 & \textbf{80.71}(+0.94\%) \\
          & ROSE  & 61.52 & 62.57 & \textit{62.61} & 62.58 & 62.37 & \textbf{64.07}(+2.55\%) \\
          & STARE & 66.15 & 66.08 & 65.54 & 66.14 & \textit{66.83} & \textbf{68.07}(+1.92\%) \\
          & ROAD  & 62.47 & 63.44 & \textit{64.93} & 64.58 & 64.25 & \textbf{65.74}(+3.27\%) \\
        \hline
    \multirow{5}[0]{*}{MD-Net} & ER    & 76.28 & \textit{76.34} &76.24 & 75.97 & 75.52 & \textbf{77.01}(+0.73\%) \\
          & MITO  & 80.28 & 80.31 & 79.97 & \textit{80.41} & 80.32 & \textbf{81.11}(+0.83\%) \\
          & ROSE  & \textit{63.31} & 63.30  & 62.77 & 62.89 & 62.90  & \textbf{65.07}(+1.76\%) \\
          & STARE & 66.46 & 66.57 & 66.43 & 67.03 & \textit{67.25} & \textbf{68.49}(+2.03\%) \\
          & ROAD  & 64.79 & 65.04 & \textit{65.23} & 65.12 & 65.15 & \textbf{66.07}(+1.28\%) \\
        \hline
    \end{tabular}
    }
    \vspace{-0.5cm}
\end{table}

\begin{table}
\caption{Ablation of step size and window size in U-Net and MD-Net.}
\label{tab4}
  \centering
 \resizebox{0.76\textwidth}{!}{
    \begin{tabular}{c|cc|cc|c|cc|cc}
    \hline
    \textbf{Model} & \textbf{Step Size}  & \textbf{Window Size} & \textbf{ER}    & \textbf{STARE} & \textbf{Model} & \textbf{Step Size}  & \textbf{Window Size} & \textbf{ER}    & \textbf{STARE} \\
    \hline
    \multirow{7}[2]{*}{U-Net*} & 1     & 11    & \textbf{76.75} & 67.87 & \multirow{7}[2]{*}{MD-Net*} & 1     & 11    & \textbf{77.03} & 68.41 \\
          & 1     & 9     & 76.73 & 67.91 &       & 1     & 9     & 77.01 & 68.21 \\
          & 1     & 7     & 76.65 & 67.89 &       & 1     & 7     & 76.92 & \textbf{68.50} \\
          & 1     & 5     & 76.59  & \textbf{68.07} &       & 1     & 5     &  77.01  & 68.49 \\
          & 2     & 5     & 76.49 & 67.87 &       & 2     & 5     & 76.91 & 68.46 \\
          & 3     & 5     & 76.68 & 67.78 &       & 3     & 5     & 76.88 & 68.32 \\
          & 4     & 5     & 76.61 & 67.78 &       & 4     & 5     & 76.90 & 68.30 \\
    \hline
    \end{tabular}%
    }
     \vspace{-0.2cm}
\end{table}

\begin{table}[t]
 \caption{Ablation study of MD-Net.}
   \label{tab5}
  \centering
   \resizebox{0.76\textwidth}{!}
   {
     \begin{tabular}{c|c|ccc|cccc|cc}
     \hline
    \textbf{Dataset} & \textbf{Model} & \textbf{FFM}   & \makecell[c]{\textbf{Edge}\\ \textbf{Decoder}} & \makecell[c]{\textbf{Skeleton}\\ \textbf{Decoder}} & \textbf{IoU$\uparrow$}   & \textbf{clDice$\uparrow$} & \textbf{ACC$\uparrow$}& \textbf{AUC$\uparrow$}   &\textbf{$\beta$  Error$\downarrow$ }    &\textbf{HD$\downarrow$}\\
     \hline
    \multirow{6}[0]{*}{ER} & U-Net &   &     &   & 75.44  & 94.63  & 91.82  & 97.35  & 28.72  & 6.87  \\
          & MD-Net &   & \CheckmarkBold    & \CheckmarkBold     & \textbf{76.28} & \textbf{95.27} & \textbf{92.00}    & \textbf{97.39} & \textbf{22.55} & \textbf{6.78} \\
              \cdashline{2-11}
          & U-Net & \CheckmarkBold &     &    & 76.59  & 95.43  & 92.02  & 97.56  & 20.78  & 6.81  \\
          & MD-Net & \CheckmarkBold    & \CheckmarkBold    &       & 76.97 & 95.60  & 92.09 & 97.59 & 22.42 & \textbf{6.71} \\
          & MD-Net & \CheckmarkBold     &       & \CheckmarkBold     & 76.85 & 95.55 & 92.13 & 97.51 & 20.63 & 6.74 \\
          & MD-Net & \CheckmarkBold    & \CheckmarkBold     & \CheckmarkBold     & \textbf{77.09} & \textbf{95.74} & \textbf{92.14} & \textbf{97.65} & \textbf{19.52} & 6.72 \\
     \hline
    \multirow{6}[0]{*}{STARE} & U-Net &     &   &    & 66.15 & 76.05 & 94.68 & 96.39 & 3.22  & 6.67 \\
          & MD-Net &      & \CheckmarkBold     & \CheckmarkBold    & \textbf{66.46} & \textbf{76.54} & \textbf{94.76} & \textbf{96.42} & \textbf{3.27} & \textbf{6.47} \\
             \cdashline{2-11}
          & U-Net & \CheckmarkBold     &      &     & 68.07 & 77.39 & 95.15 & 96.63 & 2.77  & 6.33 \\
          & MD-Net & \CheckmarkBold     & \CheckmarkBold     &       & 68.26 & 77.80  & 95.20  & 96.77 & 2.65  & \textbf{6.29} \\
          & MD-Net & \CheckmarkBold     &       & \CheckmarkBold     & 68.39 & 77.76 & 95.20  & 96.89 & 3.22  & 6.31 \\
          & MD-Net & \CheckmarkBold     & \CheckmarkBold     & \CheckmarkBold     & \textbf{68.73} & \textbf{78.20}  & \textbf{95.26} & \textbf{96.99} & \textbf{2.60}   & 6.34 \\
     \hline
    \end{tabular}
    }
    \vspace{-0.5cm}
\end{table}

\subsection{Decoders of MD-Net}
To evaluate the effectiveness of the edge decoder and the skeleton decoder, a comparative ablation analysis was conducted between MD-Net and U-Net on the ER and STARE datasets. Subsequently, we proceeded to remove either the edge decoder or the skeleton decoder from MD-Net and trained the modified models. The results shown in \cref{tab5} confirm that both the edge decoder and the skeleton decoder contribute to improving the segmentation performance. It is noteworthy that the best performance is achieved when all components are utilized simultaneously.

\subsection{Limitations of Fractal Feature Map}
 The time complexity of FFMs' computation is associated with the size of image ($M\times N$), window size $w$, and step size $S$. Its time complexity is represented as: 
$O\left(\frac{M\cdot N}{\text{S}^2} \cdot \left(w^2 \cdot \log(w) \right) \right) $. Calculation of FFM indeed incurs computational overhead but is optimized by using multithreading and larger $S$. Currently, the generation of FFMs for a $256 \times 256$ image takes about 150 milliseconds on a Xeon 8336C CPU, given a step size of 1 and a window size of 5. Detailed information on training and inference time can be found in the supplementary material.

\section{Conclusions}
In this study, we propose a method that uses fractal feature maps (FFMs) along with a multi-decoder network (MD-Net) for the semantic segmentation of tubular structures. FFMs are used to capture the texture and self-similarity of image regions or label regions at the pixel-level through the fractal dimension. $FFMs_{image}$ are used as an additional input to enhance the segmentation model's perception of tubular structures and $FFMs_{label}$ are used as a weight for the loss function to guide the model training. MD-Net improves the quality of segmentation by simultaneously predicting edges and skeletons through the incorporation of an edge decoder and a skeleton decoder. The proposed method is evaluated on five datasets of tubular structures and one dataset of non-tubular structures. The results demonstrate the superiority of MD-Net with FFM in the segmentation of tubular structures and stable performance in the segmentation of non-tubular structures. Furthermore, the performance improvement observed when incorporating the FFM module into vanilla U-Net and HR-Net underscores the broad applicability and potential of incorporating fractal information into deep learning models. 

\section*{Acknowledgements}
This work was supported in part by the National Natural Science Foundation of China (grants 92354307, 91954201, 31971289, 32101216), the Strategic Priority Research Program of the Chinese Academy of Sciences (grant XDB37040402) and the Fundamental Research Funds for the Central Universities (grant \\E3E45201X2).

\bibliographystyle{splncs04}
\bibliography{main.bbl}

\begin{thebibliography}{10}
\providecommand{\url}[1]{\texttt{#1}}
\providecommand{\urlprefix}{URL }
\providecommand{\doi}[1]{https://doi.org/#1}

\bibitem{alvarez2017tracking}
Alvarez, L., Trujillo, A., Cuenca, C., Gonz{\'a}lez, E., Esclar{\'\i}n, J., Gomez, L., Mazorra, L., Alem{\'a}n-Flores, M., Tahoces, P.G., Carreira, J.M.: Tracking the aortic lumen geometry by optimizing the 3d orientation of its cross-sections. In: Medical Image Computing and Computer-Assisted Intervention- MICCAI 2017: 20th International Conference, Quebec City, QC, Canada, September 11-13, 2017, Proceedings, Part II 20. pp. 174--181. Springer (2017)

\bibitem{antiga2003computational}
Antiga, L., Ene-Iordache, B., Remuzzi, A.: Computational geometry for patient-specific reconstruction and meshing of blood vessels from mr and ct angiography. IEEE Transactions on Medical Imaging  \textbf{22}(5),  674--684 (2003)

\bibitem{araujo2021topological}
Ara{\'u}jo, R.J., Cardoso, J.S., Oliveira, H.P.: Topological similarity index and loss function for blood vessel segmentation. arXiv preprint arXiv:2107.14531  (2021)

\bibitem{bauer2010segmentation}
Bauer, C., Pock, T., Sorantin, E., Bischof, H., Beichel, R.: Segmentation of interwoven 3d tubular tree structures utilizing shape priors and graph cuts. Medical Image Analysis  \textbf{14}(2),  172--184 (2010)

\bibitem{caicedo2019nucleus}
Caicedo, J.C., Goodman, A., Karhohs, K.W., Cimini, B.A., Ackerman, J., Haghighi, M., Heng, C., Becker, T., Doan, M., McQuin, C., et~al.: Nucleus segmentation across imaging experiments: the 2018 data science bowl. Nature Methods  \textbf{16}(12),  1247--1253 (2019)

\bibitem{caselles1997geodesic}
Caselles, V., Kimmel, R., Sapiro, G.: Geodesic active contours. International Journal of Computer Vision  \textbf{22},  61--79 (1997)

\bibitem{challoob2023distinctive}
Challoob, M., Gao, Y., Busch, A.: Distinctive phase interdependency model for retinal vasculature delineation in oct-angiography images. IEEE Transactions on Medical Imaging  (2023)

\bibitem{chen2021transunet}
Chen, J., Lu, Y., Yu, Q., Luo, X., Adeli, E., Wang, Y., Lu, L., Yuille, A.L., Zhou, Y.: Transunet: Transformers make strong encoders for medical image segmentation. arXiv preprint arXiv:2102.04306  (2021)

\bibitem{dai2017deformable}
Dai, J., Qi, H., Xiong, Y., Li, Y., Zhang, G., Hu, H., Wei, Y.: Deformable convolutional networks. In: Proceedings of the IEEE international conference on computer vision. pp. 764--773 (2017)

\bibitem{dong2022deu}
Dong, S., Pan, Z., Fu, Y., Yang, Q., Gao, Y., Yu, T., Shi, Y., Zhuo, C.: Deu-net 2.0: Enhanced deformable u-net for 3d cardiac cine mri segmentation. Medical Image Analysis  \textbf{78},  102389 (2022)

\bibitem{guo2021segmentation}
Guo, Y., Huang, J., Zhou, Y., Luo, Y., Li, W., Yang, G.: Segmentation of intracellular structures in fluorescence microscopy images by fusing low-level features. In: Pattern Recognition and Computer Vision: 4th Chinese Conference, PRCV 2021, Beijing, China, October 29--November 1, 2021, Proceedings, Part III 4. pp. 386--397. Springer (2021)

\bibitem{gupta2024topology}
Gupta, S., Zhang, Y., Hu, X., Prasanna, P., Chen, C.: Topology-aware uncertainty for image segmentation. Advances in Neural Information Processing Systems  \textbf{36} (2024)

\bibitem{he2016deep}
He, K., Zhang, X., Ren, S., Sun, J.: Deep residual learning for image recognition. In: Proceedings of the IEEE conference on computer vision and pattern recognition. pp. 770--778 (2016)

\bibitem{he2021thin}
He, Y., Ge, R., Wu, J., Coatrieux, J.L., Shu, H., Chen, Y., Yang, G., Li, S.: Thin semantics enhancement via high-frequency priori rule for thin structures segmentation. In: 2021 IEEE International Conference on Data Mining (ICDM). pp. 1096--1101. IEEE (2021)

\bibitem{hoover2003locating}
Hoover, A., Goldbaum, M.: Locating the optic nerve in a retinal image using the fuzzy convergence of the blood vessels. IEEE Transactions on Medical Imaging  \textbf{22}(8),  951--958 (2003)

\bibitem{hotamisligil2010endoplasmic}
Hotamisligil, G.S.: Endoplasmic reticulum stress and the inflammatory basis of metabolic disease. Cell  \textbf{140}(6),  900--917 (2010)

\bibitem{hu2019topology}
Hu, X., Li, F., Samaras, D., Chen, C.: Topology-preserving deep image segmentation. Advances in neural information processing systems  \textbf{32} (2019)

\bibitem{hu2021topology}
Hu, X., Wang, Y., Fuxin, L., Samaras, D., Chen, C.: Topology-aware segmentation using discrete morse theory. arXiv preprint arXiv:2103.09992  (2021)

\bibitem{huang2019batching}
Huang, Y., Tang, Z., Chen, D., Su, K., Chen, C.: Batching soft iou for training semantic segmentation networks. IEEE Signal Processing Letters  \textbf{27},  66--70 (2019)

\bibitem{isensee2021nnu}
Isensee, F., Jaeger, P.F., Kohl, S.A., Petersen, J., Maier-Hein, K.H.: nnu-net: a self-configuring method for deep learning-based biomedical image segmentation. Nature Methods  \textbf{18}(2),  203--211 (2021)

\bibitem{jin2019dunet}
Jin, Q., Meng, Z., Pham, T.D., Chen, Q., Wei, L., Su, R.: Dunet: A deformable network for retinal vessel segmentation. Knowledge-Based Systems  \textbf{178},  149--162 (2019)

\bibitem{kaplan1999extended}
Kaplan, L.M.: Extended fractal analysis for texture classification and segmentation. IEEE transactions on image processing  \textbf{8}(11),  1572--1585 (1999)

\bibitem{kaplan1994extending}
Kaplan, L.M., Kuo, C.C.: Extending self-similarity for fractional brownian motion. IEEE Transactions on Signal Processing  \textbf{42}(12),  3526--3530 (1994)

\bibitem{konatar2020box}
Konatar, I., Popovic, T., Popovic, N.: Box-counting method in python for fractal analysis of biomedical images. In: 2020 24th International Conference on Information Technology (IT). pp.~1--4. IEEE (2020)

\bibitem{kreitner2024synthetic}
Kreitner, L., Paetzold, J.C., Rauch, N., Chen, C., Hagag, A.M., Fayed, A.E., Sivaprasad, S., Rausch, S., Weichsel, J., Menze, B.H., et~al.: Synthetic optical coherence tomography angiographs for detailed retinal vessel segmentation without human annotations. IEEE Transactions on Medical Imaging  (2024)

\bibitem{la2023tubular}
La~Barbera, G., Rouet, L., Boussaid, H., Lubet, A., Kassir, R., Sarnacki, S., Gori, P., Bloch, I.: Tubular structures segmentation of pediatric abdominal-visceral cect images with renal tumors: assessment, comparison and improvement. Medical Image Analysis  \textbf{90},  102986 (2023)

\bibitem{laptev2000automatic}
Laptev, I., Mayer, H., Lindeberg, T., Eckstein, W., Steger, C., Baumgartner, A.: Automatic extraction of roads from aerial images based on scale space and snakes. Machine Vision and Applications  \textbf{12},  23--31 (2000)

\bibitem{lee2020er}
Lee, C.A., Blackstone, C.: Er morphology and endo-lysosomal crosstalk: Functions and disease implications. Biochimica et Biophysica Acta (BBA)-Molecular and Cell Biology of Lipids  \textbf{1865}(1),  158544 (2020)

\bibitem{lee2010robust}
Lee, W.L., Hsieh, K.S.: A robust algorithm for the fractal dimension of images and its applications to the classification of natural images and ultrasonic liver images. Signal Processing  \textbf{90}(6),  1894--1904 (2010)

\bibitem{li2009improved}
Li, J., Du, Q., Sun, C.: An improved box-counting method for image fractal dimension estimation. Pattern Recognition  \textbf{42}(11),  2460--2469 (2009)

\bibitem{li2023global}
Li, Y., Zhang, Y., Liu, J.Y., Wang, K., Zhang, K., Zhang, G.S., Liao, X.F., Yang, G.: Global transformer and dual local attention network via deep-shallow hierarchical feature fusion for retinal vessel segmentation. IEEE Transactions on Cybernetics  \textbf{53}(9),  5826--5839 (2023)

\bibitem{lin2013automatic}
Lin, P.L., Huang, P.W., Lee, C.H., Wu, M.T.: Automatic classification for solitary pulmonary nodule in ct image by fractal analysis based on fractional brownian motion model. Pattern Recognition  \textbf{46}(12),  3279--3287 (2013)

\bibitem{lin2015alveolar}
Lin, P., Huang, P., Huang, P., Hsu, H.: Alveolar bone-loss area localization in periodontitis radiographs based on threshold segmentation with a hybrid feature fused of intensity and the h-value of fractional brownian motion model. Computer Methods and Programs in Biomedicine  \textbf{121}(3),  117--126 (2015)

\bibitem{long2015fully}
Long, J., Shelhamer, E., Darrell, T.: Fully convolutional networks for semantic segmentation. In: Proceedings of the IEEE conference on computer vision and pattern recognition. pp. 3431--3440 (2015)

\bibitem{lopes2009fractal}
Lopes, R., Betrouni, N.: Fractal and multifractal analysis: A review. Medical Image Analysis  \textbf{13}(4),  634--649 (2009)

\bibitem{luo2020ermito}
Luo, Y., Guo, Y., Li, W., Liu, G., Yang, G.: Fluorescence microscopy image datasets for deep learning segmentation of intracellular orgenelle networks (2020). \doi{10.21227/t2he-zn97}, \url{https://dx.doi.org/10.21227/t2he-zn97}

\bibitem{ma2020rose}
Ma, Y., Hao, H., Xie, J., Fu, H., Zhang, J., Yang, J., Wang, Z., Liu, J., Zheng, Y., Zhao, Y.: Rose: a retinal oct-angiography vessel segmentation dataset and new model. IEEE Transactions on Medical Imaging  \textbf{40}(3),  928--939 (2020)

\bibitem{mnih2013machine}
Mnih, V.: Machine learning for aerial image labeling. University of Toronto (Canada) (2013)

\bibitem{mou2019cs}
Mou, L., Zhao, Y., Chen, L., Cheng, J., Gu, Z., Hao, H., Qi, H., Zheng, Y., Frangi, A., Liu, J.: Cs-net: Channel and spatial attention network for curvilinear structure segmentation. In: Medical Image Computing and Computer Assisted Intervention--MICCAI 2019: 22nd International Conference, Shenzhen, China, October 13--17, 2019, Proceedings, Part I 22. pp. 721--730. Springer (2019)

\bibitem{nain2004vessel}
Nain, D., Yezzi, A., Turk, G.: Vessel segmentation using a shape driven flow. In: Medical Image Computing and Computer-Assisted Intervention--MICCAI 2004: 7th International Conference, Saint-Malo, France, September 26-29, 2004. Proceedings, Part I 7. pp. 51--59. Springer (2004)

\bibitem{pentland1984fractal}
Pentland, A.P.: Fractal-based description of natural scenes. IEEE Transactions on Pattern Analysis and Machine Intelligence (6),  661--674 (1984)

\bibitem{qi2023dynamic}
Qi, Y., He, Y., Qi, X., Zhang, Y., Yang, G.: Dynamic snake convolution based on topological geometric constraints for tubular structure segmentation. In: Proceedings of the IEEE/CVF International Conference on Computer Vision. pp. 6070--6079 (2023)

\bibitem{roberto2021fractal}
Roberto, G.F., Lumini, A., Neves, L.A., do~Nascimento, M.Z.: Fractal neural network: A new ensemble of fractal geometry and convolutional neural networks for the classification of histology images. Expert Systems with Applications  \textbf{166},  114103 (2021)

\bibitem{ronneberger2015u}
Ronneberger, O., Fischer, P., Brox, T.: U-net: Convolutional networks for biomedical image segmentation. In: Medical Image Computing and Computer-Assisted Intervention--MICCAI 2015: 18th International Conference, Munich, Germany, October 5-9, 2015, Proceedings, Part III 18. pp. 234--241. Springer (2015)

\bibitem{shi2023affinity}
Shi, T., Ding, X., Zhou, W., Pan, F., Yan, Z., Bai, X., Yang, X.: Affinity feature strengthening for accurate, complete and robust vessel segmentation. IEEE Journal of Biomedical and Health Informatics  \textbf{27}(8),  4006--4017 (2023)

\bibitem{shit2021cldice}
Shit, S., Paetzold, J.C., Sekuboyina, A., Ezhov, I., Unger, A., Zhylka, A., Pluim, J.P., Bauer, U., Menze, B.H.: cldice-a novel topology-preserving loss function for tubular structure segmentation. In: Proceedings of the IEEE/CVF Conference on Computer Vision and Pattern Recognition. pp. 16560--16569 (2021)

\bibitem{sironi2014multiscale}
Sironi, A., Lepetit, V., Fua, P.: Multiscale centerline detection by learning a scale-space distance transform. In: Proceedings of the IEEE Conference on Computer Vision and Pattern Recognition. pp. 2697--2704 (2014)

\bibitem{taha2015metrics}
Taha, A.A., Hanbury, A.: Metrics for evaluating 3d medical image segmentation: analysis, selection, and tool. BMC Medical Imaging  \textbf{15}(1),  1--28 (2015)

\bibitem{wang2019tubular}
Wang, C., Hayashi, Y., Oda, M., Itoh, H., Kitasaka, T., Frangi, A.F., Mori, K.: Tubular structure segmentation using spatial fully connected network with radial distance loss for 3d medical images. In: Medical Image Computing and Computer Assisted Intervention--MICCAI 2019: 22nd International Conference, Shenzhen, China, October 13--17, 2019, Proceedings, Part VI 22. pp. 348--356. Springer (2019)

\bibitem{wang2020HRNet}
Wang, J., Sun, K., Cheng, T., Jiang, B., Deng, C., Zhao, Y., Liu, D., Mu, Y., Tan, M., Wang, X., et~al.: Deep high-resolution representation learning for visual recognition. IEEE Transactions on Pattern Analysis and Machine Intelligence  \textbf{43}(10),  3349--3364 (2020)

\bibitem{wang2020deep}
Wang, Y., Wei, X., Liu, F., Chen, J., Zhou, Y., Shen, W., Fishman, E.K., Yuille, A.L.: Deep distance transform for tubular structure segmentation in ct scans. In: Proceedings of the IEEE/CVF Conference on Computer Vision and Pattern Recognition. pp. 3833--3842 (2020)

\bibitem{wong2021persistent}
Wong, C.C., Vong, C.M.: Persistent homology based graph convolution network for fine-grained 3d shape segmentation. In: Proceedings of the IEEE/CVF International Conference on Computer Vision. pp. 7098--7107 (2021)

\bibitem{yan2018three}
Yan, Z., Yang, X., Cheng, K.T.: A three-stage deep learning model for accurate retinal vessel segmentation. IEEE Journal of Biomedical and Health Informatics  \textbf{23}(4),  1427--1436 (2018)

\bibitem{yang2022dcu}
Yang, X., Li, Z., Guo, Y., Zhou, D.: Dcu-net: A deformable convolutional neural network based on cascade u-net for retinal vessel segmentation. Multimedia Tools and Applications  \textbf{81}(11),  15593--15607 (2022)

\bibitem{yang2023directional}
Yang, Z., Farsiu, S.: Directional connectivity-based segmentation of medical images. In: Proceedings of the IEEE/CVF Conference on Computer Vision and Pattern Recognition. pp. 11525--11535 (2023)

\bibitem{yim2001vessel}
Yim, P.J., Cebral, J.J., Mullick, R., Marcos, H.B., Choyke, P.L.: Vessel surface reconstruction with a tubular deformable model. IEEE Transactions on Medical Imaging  \textbf{20}(12),  1411--1421 (2001)

\bibitem{yu2017dilated}
Yu, F., Koltun, V., Funkhouser, T.: Dilated residual networks. In: Proceedings of the IEEE conference on computer vision and pattern recognition. pp. 472--480 (2017)

\bibitem{zhao2017automatic}
Zhao, Y., Zheng, Y., Liu, Y., Zhao, Y., Luo, L., Yang, S., Na, T., Wang, Y., Liu, J.: Automatic 2-d/3-d vessel enhancement in multiple modality images using a weighted symmetry filter. IEEE Transactions on Medical Imaging  \textbf{37}(2),  438--450 (2017)

\bibitem{Zhou_2023_ICCV}
Zhou, Y., Huang, J., Wang, C., Song, L., Yang, G.: Xnet: Wavelet-based low and high frequency fusion networks for fully- and semi-supervised semantic segmentation of biomedical images. In: Proceedings of the IEEE/CVF International Conference on Computer Vision (ICCV). pp. 21085--21096 (October 2023)

\bibitem{zhou2018unet++}
Zhou, Z., Rahman~Siddiquee, M.M., Tajbakhsh, N., Liang, J.: Unet++: A nested u-net architecture for medical image segmentation. In: Deep Learning in Medical Image Analysis and Multimodal Learning for Clinical Decision Support: 4th International Workshop, DLMIA 2018, and 8th International Workshop, ML-CDS 2018, Held in Conjunction with MICCAI 2018, Granada, Spain, September 20, 2018, Proceedings 4. pp. 3--11. Springer (2018)

\bibitem{zhuang2019application}
Zhuang, Z., Lei, N., Joseph~Raj, A.N., Qiu, S.: Application of fractal theory and fuzzy enhancement in ultrasound image segmentation. Medical \& Biological Engineering \& Computing  \textbf{57},  623--632 (2019)

\end{thebibliography}
\end{document}